\documentclass[twocolumn]{article}
\usepackage{M-PSI} 
\removefirstpagelogo{}

\usepackage{lipsum} 
\usepackage{multirow}
\usepackage{rotating}
\usepackage{booktabs}
\usepackage{graphicx} 
\usepackage{caption}  
\usepackage{footnote} 

\usepackage{amsmath}
\usepackage{ragged2e}
\usepackage{microtype}
\usepackage{float}
\usepackage{graphicx}
\usepackage{makecell}
\usepackage{array}
\usepackage{xcolor}
\usepackage{siunitx}
\usepackage{subcaption}

\newcommand{\quotes}[1]{``#1''}

\definecolor{green}{rgb}{0, 0.7, 0}

\newcommand{\red}[1]{\textcolor{red}{#1}}
\newcommand{\blue}[1]{\textcolor{blue}{#1}}

\newcommand{\eg}{e.g., }

\usepackage[pdfencoding=auto, pdfusetitle]{hyperref}
\hypersetup{
  hidelinks,
  urlcolor=red,
  pdftitle={Exploring Dynamic Parameters for Vietnamese Gender-Independent ASR},
  pdfauthor={Sotheara Leang, Eric Castelli, Dominique Vaufreydaz, Sethserey Sam},
  pdfkeywords={speech dynamics, acoustic gesture, gender-independent automatic speech recognition, tonal and low-resource language}
}

\title{Exploring Dynamic Parameters for Vietnamese Gender-Independent ASR}
\author{Sotheara Leang\textsuperscript{\hspace{0.2mm}1,2}, \href{https://www.m-psi.fr/wp-content/uploads/2021/02/CV-CASTELLI-Eric.htm}{\'Eric Castelli\textsuperscript{\hspace{0.2mm}1,~\small{\ExternalLink}}}, \href{https://research.vaufreydaz.org/}{Dominique Vaufreydaz\textsuperscript{\hspace{0.2mm}$1$,~\small{\ExternalLink}}},
Sethserey Sam\hspace{0.2mm}$^{2}$
\vspace{0.25cm}\\
    $^1$~Univ. Grenoble Alpes, CNRS, Grenoble INP, LIG, 38000 Grenoble, France\\
    $^2$~Institute of Digital Research and Innovation, CADT, Phnom Penh, Cambodia\\
}

\begin{document}

\begin{abstract}
The dynamic characteristics of speech signal provides temporal information and play an important role in enhancing Automatic Speech Recognition (ASR). In this work, we characterized the acoustic transitions in a ratio plane of Spectral Subband Centroid Frequencies (SSCFs) using polar parameters to capture the dynamic characteristics of the speech and minimize spectral variation. These dynamic parameters were combined with Mel-Frequency Cepstral Coefficients (MFCCs) in Vietnamese ASR to capture more detailed spectral information. The SSCF0 was used as a pseudo-feature for the fundamental frequency (F0) to describe the tonal information robustly. The findings showed that the proposed parameters significantly reduce word error rates and exhibit greater gender independence than the baseline MFCCs.
\end{abstract}

\section{Introduction}
Automatic speech recognition (ASR) involves translating spoken language into written text. It maps audio signals into corresponding phonemes or characters. Recent advancements in deep learning and computing power have significantly improved ASR performance~\cite{prabhavalkar2023end}. However, it still faces challenges as speech exhibits significant variation across individuals due to their unique physiological differences, producing distinct acoustic characteristics~\cite{benzeghiba2007automatic,kent1993vocal,meunier2011vowel}. The most common acoustic features, including Mel-Frequency Cepstral Coefficient (MFCC) and filter bank features~\cite{gupta2018state,labied2021automatic,saksamudre2015review} are derived from absolute frequency measurements and exhibit a notable limitation. They are not inherently speaker-independent, often capturing speaker-specific traits such as timbre and formant patterns. Therefore, an ASR system needs extensive training data to address the diverse spectral variations, posing a significant complexity and cost challenges, especially for low-resource languages due to limited and spare datasets~\cite{besacier2014automatic}. Deep learning representations learned through self-supervised learning (\eg wav2vec) have shown significant improvements in speech recognition. However, they are not feasible and may not generalize well in low-resource scenarios~\cite{wang2024learn}.

Speech dynamics studies the temporal aspects of the speech sound, reflecting how articulatory movements, prosody, and acoustic properties shape the acoustic patterns listeners perceive. In articulatory modeling, the articulatory features describe the speech through the movements and positions of the articulators. Studies have shown that integrating these parameters with acoustic features (MFCCs, Mel filter bank energies) in ASR enhances the recognition performance and improves robustness to speaker variability~\cite{mitra2017joint,mitra2014articulatory,mitra2018articulatory}. This articulatory representation has also demonstrated greater performances in low-resource scenarios~\cite{li2019end,muller2016towards}. One drawback of this approach is that articulatory features cannot be directly extracted from the speech signal. A model must be trained to map them from the acoustic features, increasing the computational complexity.

In acoustic modeling,~\cite{carre2009signal} showed that the dynamics of vocalic transitions (movement of the formants) can be described by their direction and transition rate. In a subsequent study,~\cite{carre2017speech} conducted perception tests of the transitions from one vowel to another. The findings indicated that native listeners can identify the transition between the two vowels based on the transition direction and rate, even outside the vocalic space. This demonstrated that the dynamic parameters play a crucial role in characterizing the speech sounds and their perception.~\cite{tran2016acoustic} introduced a novel approach for characterizing vowel-to-vowel transitions by analyzing the angular transitions in Spectral Subband Centroid Frequencies (SSCFs) planes. The findings revealed that the angles are relatively independent between speaker gender and remain consistent across different speaking rates.

This research closely aligns with~\cite{carre2009signal,tran2016acoustic} and expands upon our previous study presented in~\cite{leang2022preliminary}, where we seek to characterize the acoustic transitions in the SSCF planes to describe the dynamic information of the speech signal. We hypothesize that integrating these dynamic features into ASR will enhance the ability to model diverse dynamic states and mitigate spectral variations, leading to more robust speech recognition. This study was conducted on Vietnamese, a low-resource language, focusing on speech recognition accuracy and speaker gender independence. We proposed characterizing acoustic transitions in a ratio plane of the SSCFs to enhance robustness against acoustic variations and integrating them with MFCCs to achieve better recognition. In addition, we explored the SSCF0 as a pseudo-feature for the fundamental frequency (F0) to better capture and address Vietnamese tonal characteristics.

\section{Related Work}
According to the findings from~\cite{carre2009signal},~\cite{tran2016acoustic} explored the spectral transition between vowel-to-vowel in Vietnamese using Spectral Subband Centroid Frequency (SSCFs). They possess characteristics similar to formant frequencies and are robust against noise~\cite{paliwal1998spectral}. The SSCFs can be effectively calculated without resonance or during consonant production, making them a versatile tool for analyzing various speech sounds. Six SSCFs (SSCF0 to SSCF5) were proposed to represent the fundamental frequency (F0) and the subsequent formant frequencies (F1 to F5).

\begin{equation}\label{formula_sscf}
SSCF_m= \frac{\int_{l_m}^{h_m} fw_m(f)P^\gamma(f) \,df}{\int_{l_m}^{h_m} w_m(f)P^\gamma(f) \,df}
\end{equation}

Where ${l_m}$ and ${h_m}$ represent the lower and upper bounds of subband ${m}$, ${w_m}$ denotes the subband filter, ${P(f)}$ indicates the power spectrum at frequency bin ${f}$ and ${\gamma}$ is a coefficient regulating the dynamic range of the power spectrum.\\

\begin{figure}
\center{\includegraphics[width=0.5\textwidth]{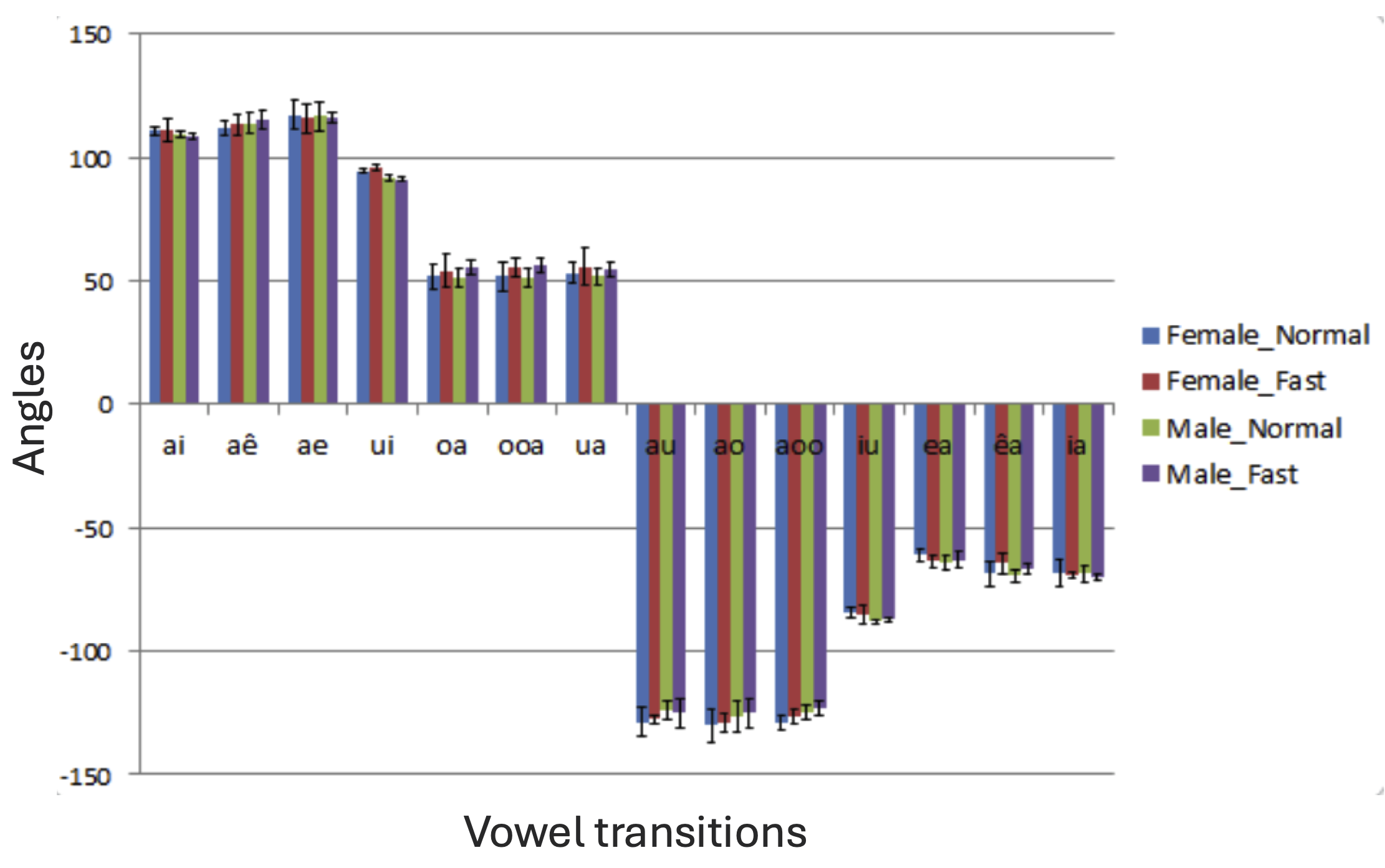}}\caption{The average angles of 14 vowel-to-vowel transitions on the SSCF1-SSCF2 plane produced by Vietnamese speakers at different speaking rates (source~\cite{tran2016acoustic}).}
\label{fig_angle_result}
\end{figure}

The SSCF trajectories of vowel-vowel transitions are relatively linear in the SSCF1-SSCF2 plane while appearing elliptical in the SSCF1-SSCF2 speed plane. This aligns with those reported by~\cite{alliot2009reconnaissance} and~\cite{carre2017speech}.~\cite{tran2016acoustic} proposed computing transition angles (see Equation \ref{formula_angle}) in each SSCF pair plane (\eg SSCF1-SSCF2 plane) to characterize the vowel transitions by assuming that the trajectories follow quasi-straight lines. The angles aim to capture the directional movement of the SSCF transitions, providing a concise representation of their dynamics. According to the study, the average angles for both male and female speakers are similar, with small standard deviation at different speaking rates for each transition (see Figure \ref{fig_angle_result}). This suggested that the angles exhibited an independence of the speaker gender and speaking rates. The combined angles, when measured in absolute values, between each transition and its corresponding reverse transition (\eg /ai/ and /ia/) are approximately 180 degrees. This suggests a symmetric relationship between the transitions, meaning the trajectories are relatively parallel.

\begin{equation}
\label{formula_angle}
Angle_{i,i+1} = \frac{\pi}{180} \arctan(\frac{\Delta SSCF_{i+1}}{\Delta SSCF_i})
\end{equation}

Where $\Delta SSCF_i$ represents the difference in $SSCF_i$ between the end and the beginning of the transition, $SSCF_i$ and $SSCF_{i+1}$ correspond to the axes of the $SSCF_i\text{-}SSCF_{i+1}$ plane.\\

\begin{figure}[ht]
\center{\includegraphics[width=0.4\textwidth]{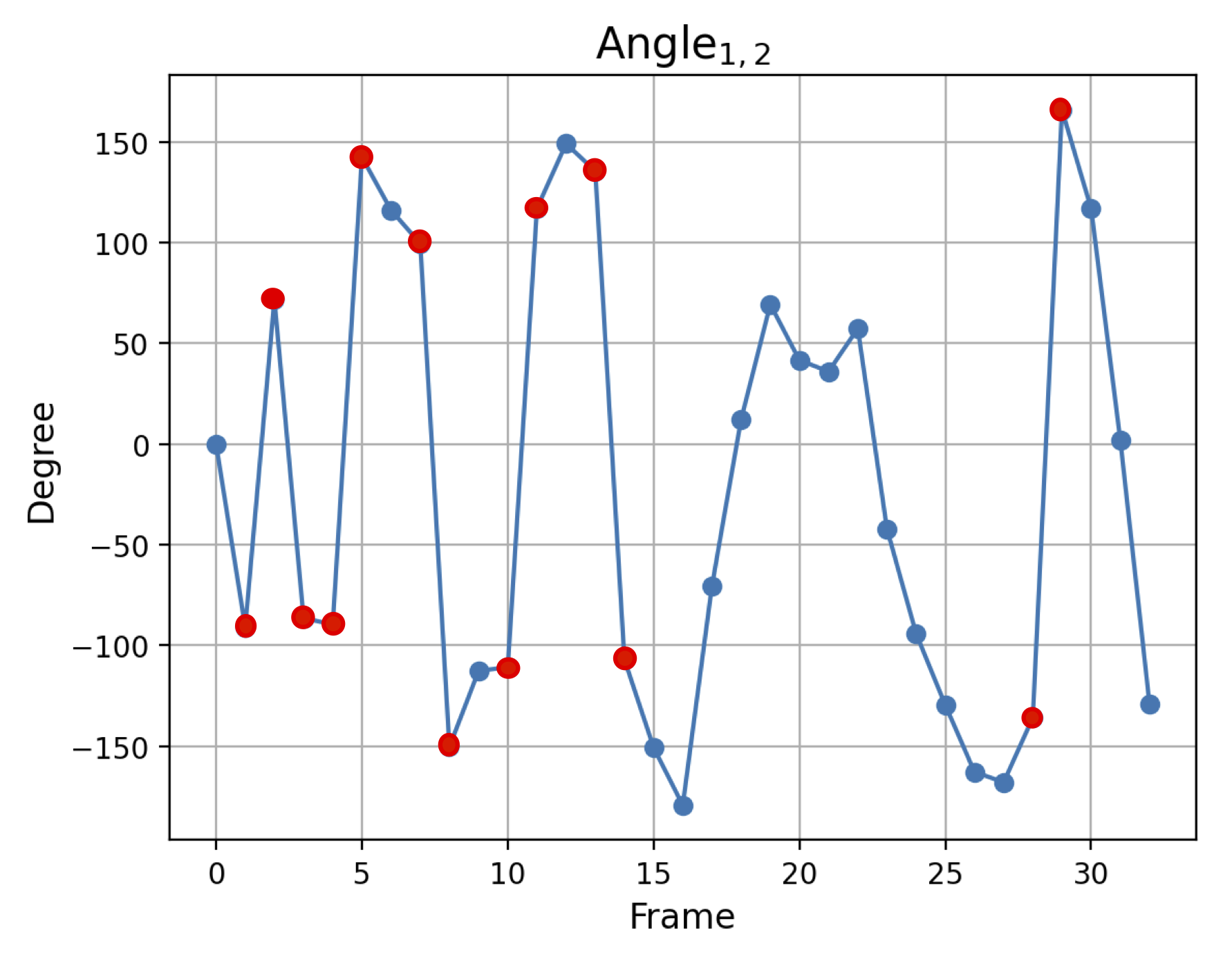}}\caption{The angles of the /ai/ transition on the SSCF1-SSCF2 plane produced by a female Vietnamese speaker. The red dots represent the start and end points of angle inversion during the transition caused by the arctan function.}
\label{fig_angle12}
\end{figure}

\begin{figure}[ht]
	\centering
	\begin{subfigure}{0.5\textwidth}\hspace{1em}
		\includegraphics[width=0.8\linewidth]{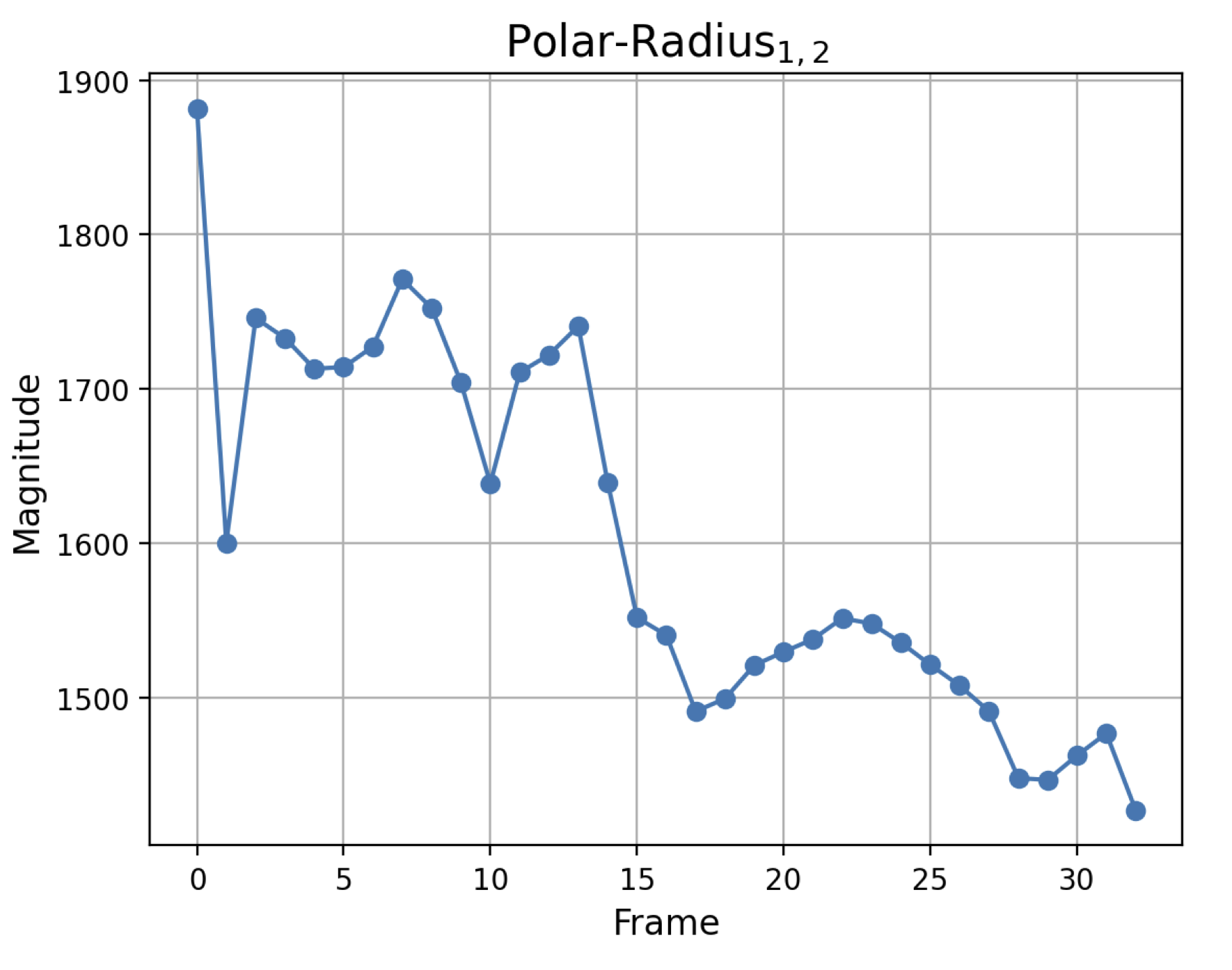}
	\end{subfigure}\\
	\begin{subfigure}{0.5\textwidth}\hspace{1em}
		\includegraphics[width=0.8\linewidth]{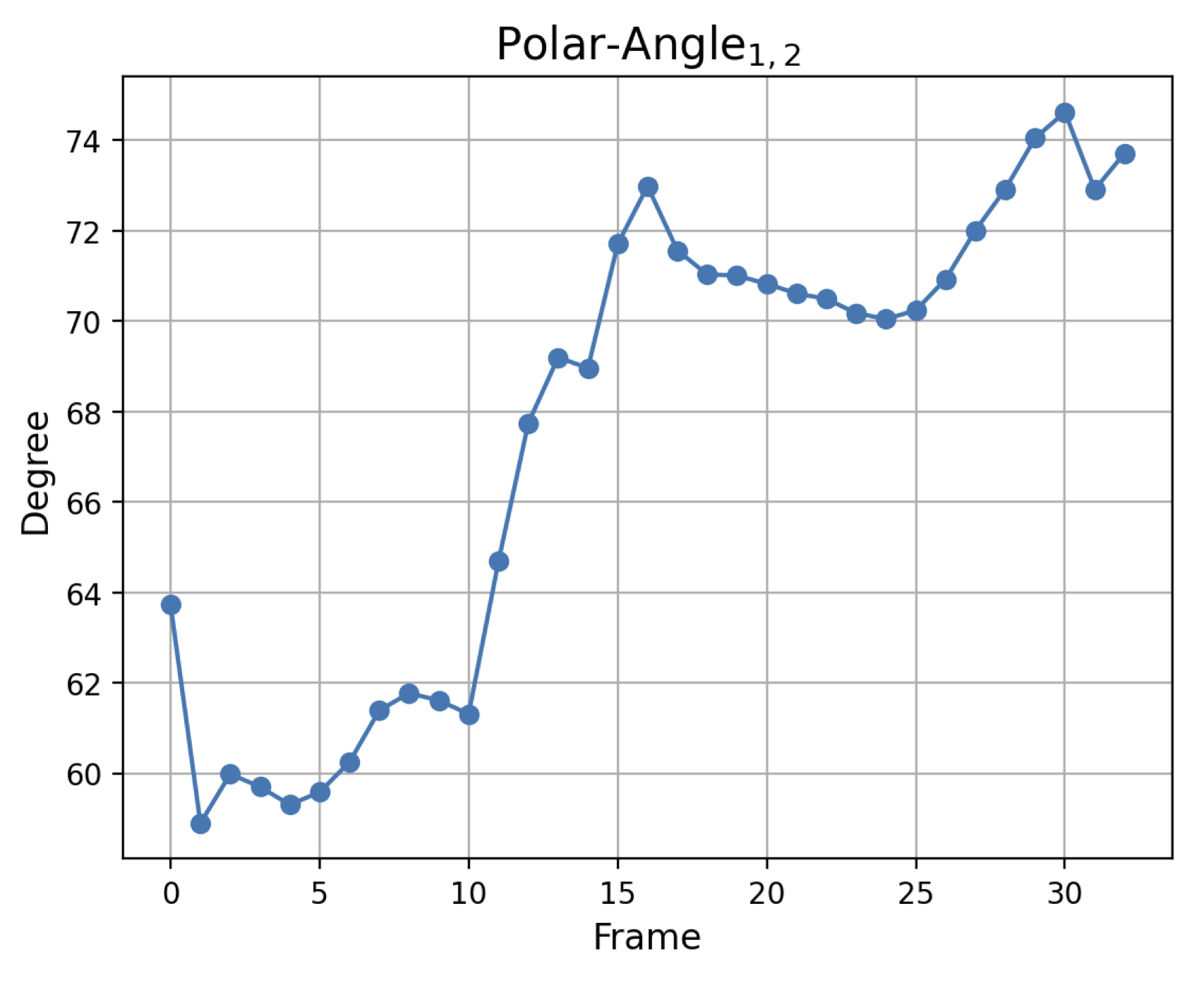}
	\end{subfigure}
	\caption{The polar parameters (radius and angle) of /ai/ transition on the SSCF1-SSCF2 plane produced by a female Vietnamese speaker.}
	\label{fig_polar_params}
\end{figure}

When applying angles in ASR, where features are extracted frame by frame, directly calculating the angles presents a significant drawback: the $\arctan$ function exhibits discontinuities at $-\pi$ and $+\pi$. This causes the inversion of the angle between positive and negative, producing noise and instability during transitions (see Figure \ref{fig_angle12}).~\cite{leang2022preliminary} proposed an alternative approach by characterizing the direction of transitions using polar coordinates. This approach showed that utilizing polar parameters, specifically radius and angle, results in a smooth and continuous trajectory, effectively capturing the spectral transitions (see Figure \ref{fig_polar_params}). The polar parameters of each SSCF pair plane were calculated using Equation \ref{formula_polar_radius} and Equation \ref{formula_polar_angle}.

\begin{equation}
\label{formula_polar_radius}
Polar\text{-}Radius_{i,i+1}(j) = \sqrt{SSCF_{i+1}(j)^2 + SSCF_{i}(j)^2} \\
\end{equation}

\begin{equation}
\label{formula_polar_angle}
Polar\text{-}Angle_{i,i+1}(j) = \frac{180^\circ}{\pi}\arctan(\frac{SSCF_{i+1}(j)}{SSCF_{i}(j)})
\end{equation}
Where $SSCF_{i}(j)$ corresponds to the $SSCF_i$ at frame j of the transition.

\section{Proposed Method}
The polar parameters proposed by~\cite{leang2022preliminary} were evaluated using a French dataset~\cite{vaufreydaz2000new} for speech recognition. The study showed that the parameters achieved higher word error rates than the MFCCs. This discrepancy arises because the polar parameters, primarily capturing dynamic states of the speech signal, omit critical detailed information essential for accurate speech recognition. Therefore, we propose combining polar parameters with the six MFCCs to capture detailed acoustic information. Only the SSCF1-SSCF2 plane was chosen to characterize the dynamic properties of the speech signal. Our study showed that other SSCF planes (SSCF2-SSCF3 plane...) do not improve the recognition results.

\subsection{Polar-Ratio Parameters}

\begin{figure}[ht]
\center{\includegraphics[width=0.4\textwidth]{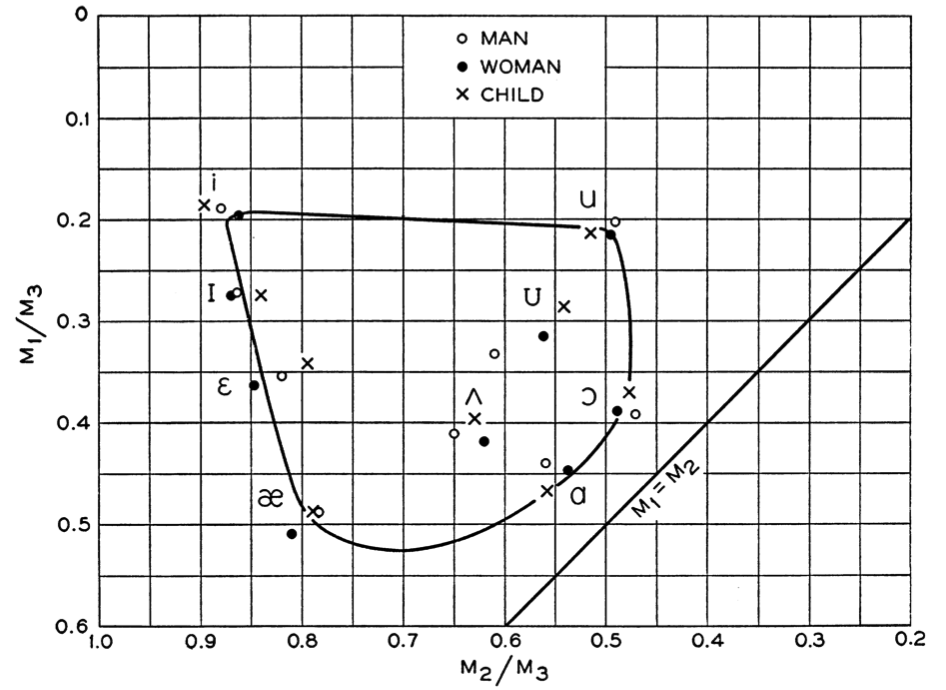}}\caption{The ratios of the first three formants (F1/F3 and F2/F3) of the vowels produced by a man, a woman and a child. The letter M denotes the formant (source~\cite{peterson1951phonetic}).}
\label{fig_formant_ratios}
\end{figure}

An acoustic analysis comparing men, women, and children was conducted in~\cite{peterson1951phonetic}. The research focused on the first two formants (F1 and F2) of the ten vowels in American English to examine the relationship between formant values and vowel identity. The findings indicated significant differences in the absolute formant values among men, women, and children due to anatomical variations. On the other hand, the study examined the formant ratios of F1/F3 and F2/F3, showing that they were relatively stable and produced a robust phonetic feature (see Figure \ref{fig_formant_ratios}). This highlights the importance of formant transitions and dynamic shifts in vowel articulation, reinforcing the role of formant ratios as an acoustic correlate of vowel perception.

According to~\cite{peterson1951phonetic}, the formant ratios (F1/F3 and F2/F3) mitigated the spectral variations between male and female speakers. Building on this finding, we propose computing the polar parameters in the ratio plane of SSCF1/SSCF3 and SSCF2/SSCF3, hypothesizing that this approach can effectively reduce gender-dependent variations in the parameters. 




\subsection{Pseudo-F0}

In this work, we deal with ASR in Vietnamese, a tonal language that relies heavily on tonal contours for phonemic distinction and accurate speech recognition. A common approach for handling tonal languages is incorporating acoustic features with the fundamental frequency (F0), as pitch information is crucial for recognizing tonal distinctions. However, accurately computing F0 in continuous speech poses significant challenges due to the inherent variability and noise in speech signals~\cite{sukhostat2015comparative}. 

To address this issue, we propose using the SSCF0 as a pseudo-F0 to provide pitch-related information for speech recognition. This approach offers a simpler method than traditional F0 computation. To account for variation between male and female speakers, the SSCF0 values were normalized using their mean and variance within the utterance. This technique has been used for normalizing acoustic features and has demonstrated that it can reduce variation in spectral features and enhance noise robustness~\cite{liu1993efficient,togneri2011overview}.

\section{Experiments}

\subsection{Datasets}
The speech corpus developed by MICA Research Institute\footnote{\url{https://mica.edu.vn/} (last seen in 04/2025)} was used in this research. Only subsets of the data were chosen as we focus on continuous speech recognition. A comparable number of utterances per speaker was chosen to create a balanced dataset suitable for cross-gender recognition. The details of the ASR corpus after preprocessing are provided in Table \ref{tbl_viet_corpus}. The corpus was further processed into three sets, each comprising train and test sets. The train sets were limited to the same size to avoid the effect of training size on the experiments. The first set, \quotes{TrainMix}, was created for general speech recognition, where the model was trained and evaluated on a dataset comprising male and female speakers. The other two sets were made for cross-gender speech recognition. In the \quotes{TrainMale}, the model was trained on male speakers, while the \quotes{TrainFemale} was trained on female speakers. Due to the small number of speakers, we proposed the experiments using 7-fold cross-validation, using 15\% of the total speakers (2 males and 2 females) as the test sets. The overall information for each cross-validation fold is given in Table \ref{tbl_viet_kfold}.

\begin{table}
\caption{The specifications of the Vietnamese corpus from MICA after preprocessing}
\centering
\begin{tabular}{|l|l|l|l|}
\hline
 \bf Hours & \bf Transcripts & \bf Speakers & \bf Vocab. \\ 
\hline 17h & 6,556 & 14 males, 14 females & 3,115  \\
\hline
\end{tabular}
\label{tbl_viet_corpus}
\end{table}

\begin{table}
\caption{The detailed datasets in each fold of the 7-fold cross-validation with the number of transcripts reported as an average.}
\centering
\begin{tabular}{|l|l|l|}
\hline
 \bf Experiment & \bf Train Set & \bf Test Sets \\ 
\hline TrainMix & \makecell{2,800 trans. \\ 6 males, 6 females} & \makecell{2 male, 500 trans. \\ 2 females, 500 trans.}  \\
\hline TrainMale & \makecell{2,800 trans. \\ 12 males \\}& \makecell{2 male, 500 trans. \\ 2 females, 500 trans.}  \\
\hline TrainFemale & \makecell{2,800 trans. \\ 12 females} & \makecell{2 male, 500 trans. \\ 2 females, 500 trans.}  \\
\hline
\end{tabular}
\label{tbl_viet_kfold}
\end{table}

\subsection{Experimental Setup}
The hybrid DNN-HMM\footnote{DNN-HMM: Deep Neural Network - Hidden Markov Model}-based ASR system\footnote{\url{https://github.com/kaldi-asr/kaldi/blob/master/egs/wsj/s5/local/nnet2/run\_5c.sh  (last seen in 04/2025)}} from the Kaldi toolkit~\cite{ghoshal2011kaldi} was adopted as the foundation tool for this study. In line with our previous work~\cite{leang2022preliminary}, the context-independent model was built with 1,000 Gaussians and trained for 40 iterations. The context-dependent model consisted of 2,000 states and 10,000 Gaussians, and was trained for 35 iterations. The 3gram-language model was trained with SRILM using the complete transcripts of the corpus to achieve a good language model, as we mainly evaluate the performance of the acoustic model. The lexicon developed by MICA was utilized, with tones treated as distinct features, similarly to phonemes, and considered at the same level.

The DNN model processes nine consecutive contextual frames as input, passing through three hidden layers, each containing 512 neurons. The hyperbolic tangent function serves as the activation function for all neurons. The neural training was conducted for 20 epochs with a batch size of 128, using an initial learning rate of 0.01, gradually decreasing to a final learning rate of 0.001.

The MFCCs with 6 and 13 dimensions were used as baseline features, derived from 6 and 13 subband filters, respectively. Feature extraction for all parameters, including the proposed features, was conducted using a 25-millisecond window length and a 10-millisecond hop length. During the process, a 0.97 pre-emphasis factor and a Hamming window were applied, and the spectrum was computed using an  FFT\footnote{\url{FFT: Fast Fourier Transform}} size of 512. A cepstral liftering coefficient of 22 was used for MFCC feature extraction.

\section{Results and Discussions}
The recognition results in Table \ref{tbl_results} indicate that the 13 MFCCs performed significantly better than the 6 MFCCs in TrainMix. This enhancement is due to the greater resolution offered by the 13 filters compared to the 6 filters. They also achieved better results in TrainFemale, except when evaluated on female speech in TrainMale. This suggests that training on female speech generally results in lower error rates than male speech, as female speech tends to exhibit greater acoustic variability, enhancing model generalization. However, the 6 MFCCs exhibited more gender independence (smaller |M-F| in TrainMale and TrainFemale), indicating that higher-resolution features may hinder cross-gender speech recognitions by capturing more gender-specific details, which reduces the ability to generalize across genders.

When the 6 MFCCs were combined with the polar parameters, the error rates were significantly reduced across all experiments, and they surpassed the 13 MFCCs in cross-gender speech recognitions while achieving comparable performance in TrainMix. This demonstrates that the dynamic features provided by the polar parameters capture more detailed information essential for speech recognition, thereby enhancing performance. Additionally, applying the polar ratio resulted in marginal improvements in various cases, and the parameters demonstrated greater gender independence.

Incorporating the SSCF0 resulted in slight improvements in certain cases. The best performance was observed when mean and variance normalization (MVN) was applied, as it reduced the spectral dispersion of the SSCF0 between speakers, making the parameters greater independent. These findings confirm that SSCF0 enhances Vietnamese speech recognition by capturing valuable information related to the fundamental frequency (F0). Overall, the proposed method achieves superior recognition results, provides more speaker gender independence, and outperforms the 13 MFCCs while utilizing fewer parameters.

\begin{table*}[ht]
\caption{Word error rates (\%) comparing the proposed parameters with baseline features using 6 and 13 MFCCs. The best and least favorable results are highlighted in blue and red. $\Delta$ and $\Delta \Delta$ denote the first- and second-order derivative features. MVN stands for mean and variance normalization. |M-F| represents the difference in word error rates between test males and females.}
\centering
\begin{tabular}{|l|l|l|l|l|l|l|}
\hline
\textbf{Parameters} & \multicolumn{2}{|c|}{\textbf{TrainMix}} & \multicolumn{2}{|c|}{\textbf{TrainMale}} & \multicolumn{2}{|c|}{\textbf{TrainFemale}} \\ \hline

6 MFCC, $\Delta$, $\Delta\Delta$ 
& \begin{tabular}[c]{@{}c@{}}M \\ F \\ All \\ \textbar M-F \textbar \end{tabular}
& \begin{tabular}[c]{@{}S[table-format=2.2]@{}} \red{14.17} \\ \red{10.58} \\ \red{ 12.77} \\ \red{\hphantom{0}3.59} \end{tabular}
& \begin{tabular}[c]{@{}c@{}}M \\ F \\ All \\ \textbar M-F \textbar\end{tabular}
& \begin{tabular}[c]{@{}S[table-format=2.2]@{}} \red{13.04} \\ 14.56 \\ \red{14.06} \\ \blue{\hphantom{0}1.52} \end{tabular} 
& \begin{tabular}[c]{@{}c@{}}M \\ F \\ All \\ \textbar M-F \textbar\end{tabular} 
& \begin{tabular}[c]{@{}S[table-format=2.2]@{}} \red{ 15.49} \\ \red{11.51} \\ \red{14.14} \\ \red{\hphantom{0}3.98} \end{tabular} \\ \hline

13 MFCC, $\Delta$, $\Delta \Delta$ 
& \begin{tabular}[c]{@{}c@{}}M \\ F \\ All \\ \textbar M-F \textbar \end{tabular}
& \begin{tabular}[c]{@{}S[table-format=2.2]@{}} 10.17 \\ \hphantom{0}\blue{8.19} \\ \blue{ \hphantom{0}9.46} \\ 1.98 \end{tabular}
& \begin{tabular}[c]{@{}c@{}}M \\ F \\ All \\ \textbar M-F \textbar\end{tabular}
& \begin{tabular}[c]{@{}S[table-format=2.2]@{}} 9.41 \\ \red{ 16.00} \\ 12.72 \\ \red{\hphantom{0}6.59} \end{tabular} 
& \begin{tabular}[c]{@{}c@{}}M \\ F \\ All \\ \textbar M-F \textbar\end{tabular} 
& \begin{tabular}[c]{@{}S[table-format=2.2]@{}} 13.36 \\ 9.75 \\ 12.15 \\ 3.61 \end{tabular} \\ \hline
\Xhline{1.5\arrayrulewidth}

\makecell[l]{(6 MFCC, $\Delta$, $\Delta \Delta$), \\ Polar} 
& \begin{tabular}[c]{@{}c@{}}M \\ F \\ All \\ \textbar M-F \textbar \end{tabular}
& \begin{tabular}[c]{@{}S[table-format=2.2]@{}} 10.30 \\ 8.98 \\ 9.89 \\ \blue{\hphantom{0}1.32} \end{tabular}
& \begin{tabular}[c]{@{}c@{}}M \\ F \\ All \\ \textbar M-F \textbar\end{tabular}
& \begin{tabular}[c]{@{}S[table-format=2.2]@{}} 9.39 \\ 14.17 \\ 11.78 \\ 04.78 \end{tabular} 
& \begin{tabular}[c]{@{}c@{}}M \\ F \\ All \\ \textbar M-F \textbar\end{tabular} 
& \begin{tabular}[c]{@{}S[table-format=2.2]@{}} 12.68 \\ 9.80 \\ 11.77 \\ 2.88 \end{tabular} \\ \hline

\makecell[l]{(6 MFCC, $\Delta$, $\Delta \Delta$), \\ Polar-Ratio} 
& \begin{tabular}[c]{@{}c@{}}M \\ F \\ All \\ \textbar M-F \textbar \end{tabular}
& \begin{tabular}[c]{@{}S[table-format=2.2]@{}} 10.44 \\ 8.91 \\ 9.94 \\ 1.53 \end{tabular}
& \begin{tabular}[c]{@{}c@{}}M \\ F \\ All \\ \textbar M-F \textbar\end{tabular}
& \begin{tabular}[c]{@{}S[table-format=2.2]@{}} 9.77 \\ 13.23 \\ 11.64 \\ 3.46 \end{tabular} 
& \begin{tabular}[c]{@{}c@{}}M \\ F \\ All \\ \textbar M-F \textbar\end{tabular} 
& \begin{tabular}[c]{@{}S[table-format=2.2]@{}} 12.28 \\ 9.81 \\ 11.53 \\ 2.47 \end{tabular} \\ \hline

\makecell[l]{(6 MFCC, $\Delta$, $\Delta \Delta$), \\ Polar-Ratio, SSCF0} 
& \begin{tabular}[c]{@{}c@{}}M \\ F \\ All \\ \textbar M-F \textbar \end{tabular}
& \begin{tabular}[c]{@{}S[table-format=2.2]@{}} 10.27 \\ 8.78 \\ 9.79 \\ 1.49 \end{tabular}
& \begin{tabular}[c]{@{}c@{}}M \\ F \\ All \\ \textbar M-F \textbar\end{tabular}
& \begin{tabular}[c]{@{}S[table-format=2.2]@{}} 9.50 \\ 13.05 \\ 11.38 \\ 3.55 \end{tabular} 
& \begin{tabular}[c]{@{}c@{}}M \\ F \\ All \\ \textbar M-F \textbar\end{tabular} 
& \begin{tabular}[c]{@{}S[table-format=2.2]@{}} 11.72 \\ 9.72 \\ 11.14 \\ \blue{\hphantom{0}2.00} \end{tabular} \\ \hline

\makecell[l]{(6 MFCC, $\Delta$, $\Delta \Delta$), \\ Polar-Ratio, SSCF0-MVN} 
& \begin{tabular}[c]{@{}c@{}}M \\ F \\ All \\ \textbar M-F \textbar \end{tabular}
& \begin{tabular}[c]{@{}S[table-format=2.2]@{}} \blue{10.02} \\ 8.47 \\ 9.50 \\ 1.55 \end{tabular}
& \begin{tabular}[c]{@{}c@{}}M \\ F \\ All \\ \textbar M-F \textbar\end{tabular}
& \begin{tabular}[c]{@{}S[table-format=2.2]@{}} \hphantom{0}\blue{9.32} \\ \blue{ 12.57} \\ \blue{11.05} \\ 3.25 \end{tabular} 
& \begin{tabular}[c]{@{}c@{}}M \\ F \\ All \\ \textbar M-F \textbar\end{tabular} 
& \begin{tabular}[c]{@{}S[table-format=2.2]@{}} \blue{ 11.69} \\ \blue{\hphantom{0}9.48} \\ \blue{11.01} \\ 2.21 \end{tabular} \\ \hline

\end{tabular}
\label{tbl_results}
\end{table*}

\section{Conclusion}
This research extended the previous study on exploring the dynamic parameters by characterizing the acoustic transitions in the SSCF plane. We proposed to use polar parameters on the ratio plane of SSCF1/SSCF3 and SSCF2/SSCF3. This approach was inspired by~\cite{peterson1951phonetic}, which explored the formant ratios (F1/F3, F2/F3), asserting that such ratios can help reduce spectral variation. Relying solely on polar parameters may not capture sufficient detail for effective speech recognition. Therefore, we proposed combining the polar ratio parameters with the MFCC features. The proposed method was applied to the Vietnamese data, and the results showed that the proposed parameters significantly outperformed and were more gender-independent. Additionally, the SSCF0 was introduced as a pseudo-F0 to address tonal information to the Vietnamese. By incorporating the normalized SSCF0 into the proposed parameters, we achieved lower word error rates while maintaining greater gender independence than the baseline MFCCs.

In future work, we seek to evaluate the proposed method on additional languages such as French and Khmer and to expand the analysis to larger datasets with more speakers to validate its generalization. In addition, applying the proposed parameters to more advanced end-to-end neural models would be valuable for investigating their impact. Lastly, we intend to explore the characterization of the transition rate, taking inspiration from~\cite{carre2009signal} as the current method only describes the direction of the acoustic transitions. Incorporating this additional information could provide more detailed information on the dynamic aspects of the speech signal, improving  speech recognition accuracy.

\section*{Acknowledgments}
This research was supported by French Government Scholarship (BGF) from the French Embassy in Cambodia and Cambodia Academy of Digital Technology (CADT). This research was made possible by the collaboration between M-PSI team at Grenoble Informatics Laboratory (LIG) and CADT.

\bibliographystyle{plainnat}
\bibliography{biblio}

\end{document}